\pdfoutput=1
\documentclass[11pt]{arxiv-article}
  
\bibliography{References}

\usepackage{Definitions}
\usepackage{amssymb,amsfonts,multirow}

\usepackage{pifont}
\usepackage{hyperref}
\usepackage{mathrsfs}
\usepackage{mathdots}
\makeatletter
\renewcommand*\env@matrix[1][*\c@MaxMatrixCols c]{%
  \hskip -\arraycolsep
  \let\@ifnextchar\new@ifnextchar
  \array{#1}}
\makeatother

\newcommand{\OfficialTitle}{
  A safe CFT at large charge
}
\title{\setstretch{1.4}
  {\color{Thoughtless}\Huge\textbf{\dosserif\OfficialTitle}}
}

\hypersetup{pdfauthor={Domenico Orlando and Susanne Reffert and Francesco Sannino},pdftitle={\OfficialTitle},%
  colorlinks=true,linkcolor=ThoughtYouWere,citecolor=ThoughtYouWere,urlcolor=ThoughtYouWere}

\author{%
  \begin{minipage}{.97\linewidth}
    \vspace{1cm}
    \begin{center} \dosserif%
      {\small
         \textbf{Domenico~Orlando}\textsuperscript{\ding{71}\ding{72}},
         \textbf{Susanne~Reffert}\textsuperscript{\ding{72}} and
         \textbf{Francesco~Sannino}\textsuperscript{\ding{73}} 
         }
    \end{center}
    \vspace{1cm}
     \authorBlock{\ding{71}}{\dosserif{} INFN sezione di Torino | Arnold--Regge Center\\
      via Pietro Giuria 1, 10125 Torino, Italy}
    \authorBlock{\ding{72}}{\dosserif{} Albert Einstein Center for Fundamental Physics\\
     Institute for Theoretical Physics, University of Bern,\\
     Sidlerstrasse 5, CH-3012 Bern, Switzerland}
     \authorBlock{\ding{73}}{\dosserif{}CP3-Origins \& the Danish Institute for Advanced Study\\
University of Southern Denmark\\ Campusvej 55, DK-5230 Odense, Denmark}
  \end{minipage}
}

\date{}

\begin{document}

\setstretch{1.2}

\numberwithin{equation}{section}

\begin{titlepage}

  \newgeometry{top=23.1mm,bottom=46.1mm,left=34.6mm,right=34.6mm}

  \maketitle

  \thispagestyle{empty}

  \vfill\dosserif{}

  \abstract{\normalfont{}\noindent{}%
    We apply the large-charge limit to the first known example of a four-dimensional gauge-Yukawa theory featuring an ultraviolet interacting fixed point in all couplings. %
     We determine the energy of the ground state in presence of large fixed global charges and deduce the global symmetry breaking pattern. We show that the fermions decouple at low energy leaving behind a confining Yang--Mills theory and a characteristic spectrum of type~I (relativistic) and type~II (non-relativistic) Goldstone bosons. Armed with the knowledge acquired above we finally arrive at establishing the conformal dimensions of the theory as a triple expansion in the large-charge, the number of flavors and the controllably small inverse gauge coupling constant at the UV fixed point. Our results unveil a number of noteworthy properties of the low-energy spectrum, vacuum energy and conformal properties of the theory. They also allow us to derive a new consistency condition for the relative sizes of the couplings at the fixed point. %
  }

\vfill

\end{titlepage}

\restoregeometry{}

\setstretch{1.2}

\tableofcontents

\section{Introduction}%
\label{sec:into}

Gauge--Yukawa theories constitute the backbone of the standard model of particle interactions and of most of its extensions. It is therefore paramount to elucidate their dynamics starting from those theories that are fundamental according to Wilson~\cite{Wilson:1971bg,Wilson:1971dh}. This means that {they} are well-defined at arbitrarily short distances. Asymptotically free~\cite{Gross:1973ju,Politzer:1973fx}  and safe~\cite{Litim:2014uca} quantum field theories are distinct types of fundamental quantum field theories. Freedom implies that at extremely short distances all interactions vanish while safety requires a freezing of the interactions.    Although asymptotic freedom has a long history, the discovery of four-dimensional controllable asymptotically safe quantum field theories is quite recent~\cite{Litim:2014uca,Litim:2015iea}. Any fundamental field theory is governed at short distances by a \ac{cft} which is interacting in the safe case.  It is therefore timely to explore safe \ac{cft} dynamics starting with the first example of such a theory introduced in~\cite{Litim:2014uca}.   

Here we shall be concerned with generalizing and applying the large-charge limit of \acp{cft} to four-dimensional controllable nonsupersymmetric asymptotically safe quantum field theories~\cite{Litim:2014uca,Litim:2015iea}. This limit has proven useful, in the past, when investigating generic properties of \acp{cft} with global symmetries.
Restricting our attention to a subsector of the theory of large fixed charge leads to important simplifications.
Fixing the charge introduces a scale into the otherwise scale-free problem -- the charge density $\rho$.
It plays the role of a controlling parameter in a perturbative expansion.
This expansion works in the energy range
\begin{equation}\label{eq:hierarchy}
	\Lambda^Q_{\text{low}}=\frac{1}{r_0} \ll \Lambda \ll \Lambda^Q_{\text{high}}=\rho^{1/d}
\end{equation}
for a theory in $d+1$ dimensions compactified on a $d$-dimensional manifold of characteristic scale $r_0$.
In analogy to chiral perturbation theory, the natural expansion parameter is given by
\begin{equation}
	\frac{\Lambda_{\text{low}}^Q}{\Lambda_{\text{high}}^Q}=\frac{1}{Q^{1/d}}.
\end{equation}
Solving the classical \ac{eom} at fixed charge gives rise to a time-dependent ground state. Studying the fluctuations around this fixed-charge ground state is formally equivalent to studying fluctuations at fixed chemical potential. Working at fixed charge ``classicalizes'' the problem, in the sense that the largest contributions (with positive scaling in $Q$) come from the classical solution, while the quantum corrections are controlled by inverse powers of the large charge.

The breaking of the conformal and global symmetries by the ground state gives rise to a set of both relativistic and non-relativistic Goldstone fields which encode the low-energy dynamics. It is in fact possible to write a non-linear sigma model for the Goldstones, in which the inverse charge controls both the tree-level contributions and the quantum effects.  The leading quantum correction is given by the Casimir-energy of the relativistic Goldstone bosons alone, as the non-relativistic Goldstone bosons do not have a zero-point energy. Among the relativistic Goldstones, there is always one special mode (the \emph{conformal Goldstone}) with a speed of sound equal to $1/\sqrt{d}$ which is a remnant of the broken conformal invariance. 

The large-charge expansion allows us to calculate \ac{cft} data. Generally speaking, conformal dimensions \(\Delta\) of operators on \(\setR^{d+1}\) are related to the energies \(E_{S^d}\) of states living on a $d$-sphere of radius \(r_0\) through the relation \(\Delta = r_0 E_{S^d}\)~\cite{Cardy:1984rp,Cardy:1985lth}.
This relation, known as the state-operator correspondence, is a consequence of the fact that \(\setR^{d+1}\) is conformally equivalent to \(\setR \times S^d(r_0)\). The state-operator correspondence allows us in particular to directly calculate the conformal dimension of the lowest-lying operator of charge $Q$ by evaluating the energy of the ground state on the $d$-sphere.

So far, the large-charge expansion has been used mostly for three-dimensional \acp{cft} at an \ac{ir} interacting fixed point (see e.g.~\cite{Hellerman:2015nra,Alvarez-Gaume:2016vff,Monin:2016jmo,Loukas:2017lof,Watanabe:2019adh}). The large-charge results for the $O(N)$ vector model have been confirmed to high precision for $N=2$ and $N=4$ on the lattice~\cite{Banerjee:2017fcx,Banerjee:2019jpw}.
Here it was discovered  that in these models the large-charge prediction for the conformal dimension holds also for order one charges. In four dimensions the method has been applied only to either supersymmetric~\cite{Hellerman:2017sur,Hellerman:2017veg,Hellerman:2018xpi,Beccaria:2018xxl,Bourget:2018obm} theories or non-relativistic~\cite{Favrod:2018xov,Kravec:2018qnu,Kravec:2019djc} examples.

We find it therefore highly interesting to move into the four-dimensional nonsupersymmetric realm and report our central result for  our    safe   \ac{cft}:
\begin{multline}
  \Delta(J ) = r_0 E(S^3) = \frac{3}{2} \frac{N_F^2}{\alpha_h + \alpha_v}  \bqty{ \Jexp^{4/3} + \frac{1}{6} \Jexp^{2/3} - \frac{1}{144}  \Jexp^0 + \order{\Jexp^{-2/3}}} \\ - \pqty{ \pqty{ \frac{N_F^2}{2}  - 2} \sqrt{\frac{\alpha_h}{3 \alpha_h + 2 \alpha_v}} + \frac{1}{\sqrt{3}}} \times 0.212\dots + \order{\Jexp^{-2/3}},
\end{multline}
where $\alpha_h$ and $\alpha_v$ are the opportunely normalized scalar self-couplings at the \ac{uv} fixed point, $N_f$ is the number of flavors of the theory and ${\cal J}\gg 1$ the opportunely normalized charge. The last term derives from the vacuum energy of the relativistic Goldstones which is of order ${\cal J}^0$. As expected, the conformal Goldstone contributes a factor $1/\sqrt{3}$ in four dimensions.

\medskip
 The paper is organized as follows: In Section~\refstring{sec:model} we introduce the model of~\cite{Litim:2014uca}. Section~\refstring{sec:scalar-sector} deals with introducing the fixed charges, solving the equations of motion, determining the ground state and elucidating the pattern of symmetry breaking. In Section~\refstring{sec:fermions} we discuss the fermion decoupling and the emergence of a decoupled low energy Yang--Mills theory. The Goldstone spectrum and its properties are derived in Section~\refstring{sec:Goldstone-spectrum-equal}. Here we also determine the critical quantities. Finally, we offer our conclusions in Section~\refstring{sec:conclusions}. In Appendix~\ref{sec:general} we consider a  general setting for the choice of fixed charges.

\section{The Model}%
\label{sec:model}

We start from a \ac{cft} in four dimensions, containing $SU(N_C)$ gauge fields $A_\mu^a$, $N_F$ flavors of fermions $Q_i$ in the fundamental and an $N_F \times N_F$ complex matrix scalar field $H$ which is not charged under $SU(N_C)$. In the Veneziano limit of $N_F \to \infty, \ N_C \to \infty$ with the ratio $N_F/N_C$ fixed, this theory is asymptotically safe, as shown in~\cite{Litim:2014uca}.
Its Lagrangian is given by
\begin{equation}\label{eq:fullLag}
  \begin{aligned}
    \mathcal{L} ={}& -\frac{1}{2}\Tr (F^{\mu\nu}F_{\mu\nu}) + \Tr(\bar Q i\slashed{D} Q) + y \Tr(\bar Q_L H Q_R + \bar Q_R H^\dagger Q_L)\\
    &+ \Tr(\del_\mu H^\dagger \del^\mu H ) - u\Tr(H^\dagger H)^2 - v(\Tr H^\dagger H )^2- \frac{R}{6} \Tr( H^\dagger H) .
  \end{aligned}
\end{equation}
The trace runs over both color and flavor indices and $Q_{L/R} = \tfrac{1}{2} (1\pm \gamma_5)Q$.
In view of the state/operator correspondence we have added a conformal coupling term for the bosonic fields, proportional to the Ricci scalar $R$.
No such term is needed for the fermions or the gauge fields in four dimensions.
From now on, we will consider our theory on \(\setR \times M_3\), where \(M_3\) is a compact manifold (eventually the three-sphere when computing the conformal dimensions).

The rescaled couplings of the model appropriate for the Veneziano limit are
\begin{align}\label{eq:couplings}
  \alpha_g & = \frac{g^2 N_C}{(4\pi)^2}, & \alpha_y & = \frac{y^2 N_C}{(4\pi)^2}, & \alpha_h & = \frac{u N_F}{(4\pi)^2}, & \alpha_v & = \frac{v N_F^2}{(4\pi)^2},
\end{align}
where $\alpha_g$ is the gauge coupling (as opposed to the original gauge coupling \(g\) in Eq.~\eqref{eq:fullLag}), $\alpha_y$ the Yukawa coupling, $\alpha_h$ the quartic scalar coupling and $\alpha_v$ the double-trace coupling.

We also introduce the control parameter 
\begin{equation}
	\epsilon = \frac{N_F}{N_C}-\frac{11}{2},
\end{equation}
which in the Veneziano limit is continuous and arbitrarily small.
The action in Eq.~\eqref{eq:fullLag} has \(U(N_F) \times U(N_F)\) symmetry, but we will concentrate on the quantum symmetry  $SU(N_F)_L \times SU(N_F)_R \times U(1)_B$ since we know that there is an anomalous axial \(U(1)\).

As shown in Ref.~\cite{Litim:2014uca}, if \(0 \leq \epsilon \ll 1\), within the perturbative regime there is one fixed point that is unique in that it has only one relevant direction with the other three being irrelevant.

To the maximum currently achievable order in perturbation theory and properly respecting the Weyl consistency conditions~\cite{Antipin:2013sga} it is obtained for 
\begin{equation}
  \label{NNLOseries}
  \begin{aligned}
\alpha_g^*&= 0.4561\,\epsilon+0.7808 \,\epsilon^2
+\order{\epsilon^3}\\[.5ex]
\alpha_y^*&= 0.2105\,\epsilon+0.5082\,\epsilon^2
+\order{\epsilon^3}\\[.5ex]
\alpha_h^*&= 0.1998\,\epsilon+0.5042\,\epsilon^2
+\order{\epsilon^3}\,,
\end{aligned}
\end{equation}
with the leading coefficients of $\epsilon$ corresponding to $\alpha_g^* = \frac{26}{57}\epsilon +\ldots$, $\alpha_y^* = \frac{4}{19}\epsilon +\ldots$ and $\alpha_h^* = \frac{\sqrt{23}-1}{19}\epsilon +\ldots$ respectively. Note that the Yukawa and quartic scalar self-couplings are essential for this fixed point to exist. The remaining double-trace scalar coupling $v$ has two possible fixed points, one of which is reliable and adds an irrelevant scaling direction to the theory, for 
\begin{equation}
\label{v1}
\alpha_{v1}^* =
\frac{-6 \sqrt{23+4 \epsilon} + 3\sqrt{4 \epsilon +6 \sqrt{23+4 \epsilon}+20}}{4 \epsilon +26}\alpha_g^*
+ \order{{\alpha_g^*}^2} .
\end{equation}
Numerically, $\alpha_{v1} ^*=-0.1373\,\epsilon$ up to quadratic corrections in $\epsilon$.

\medskip
We now want to apply the large-charge expansion~\cite{Hellerman:2015nra} to this theory, which will lead to an effective perturbative action in the Goldstone \acp{dof} resulting from the symmetry breaking induced by the fixed-charge ground state, where higher-order terms will be suppressed by inverse powers of the fixed charge.
We can use this action to calculate the anomalous dimension of the lowest operator of fixed charge.
We will be working at a scale $\Lambda$, where
\begin{equation}
  \frac{1}{r_0} \ll \Lambda_{UV} \ll \Lambda \ll \rho^{1/3},
\end{equation}
where $r_0$ is the typical scale of our compactification manifold, $\rho$ is the charge density of the fixed charges as in Eq.~\eqref{eq:hierarchy}, and $\Lambda_{UV}$ is the scale of the \ac{uv} fixed point at which the couplings in Eq.~\eqref{NNLOseries} are evaluated. Fixing the charge effectively introduces a new relevant direction which drives us away from the \ac{uv} fixed point to a new \ac{ir} fixed point.

  The hierarchy of scales is chosen so that we are at the \ac{uv} point where the effective action is consistent and that the scale corresponding to fixing the charge is the dominating one.

In the following we will always consider this fixed point.

To make the notation lighter we will use the unstarred couplings  but we will intend them to be evaluated at this fixed point. All the expressions are to be understood up to order \(\order{\epsilon} \) corrections.

\section{Fixing the charges in the scalar sector}
\label{sec:scalar-sector}

\subsection{Equations of motion and ground state}
\label{sec:EOM-ground-state}

We will first focus on the sector involving just the scalar field $H$, using the Lagrangian
\begin{equation}\label{eq:scalarL}
  \mathcal{L}_H = \Tr(\del_\mu H^\dagger \del^\mu H ) - u\Tr(H^\dagger H)^2 - v(\Tr H^\dagger H )^2 - \frac{R}{6} \Tr( H^\dagger H). 
\end{equation}
In a later step, we will show that indeed, all the fermions $Q_i$ will receive large masses from fixing the charge and, together with the gluons, decouple from the dynamics. We will follow the procedure outlined in~\cite{Loukas:2017lof} for fixing the charge in matrix models.

When focusing only on the scalar sector, the model naively has a global \(U(N_F)_L \times U(N_F)_R\) symmetry at the classical level. We know however that the full model (given in Eq.~\eqref{eq:fullLag}) has an axial anomaly due to the Yukawa term. For this reason, we work directly in the quantum global symmetry group \(SU(N_F)_L \times SU(N_F)_R \times U(1)_B\) which is generated by the currents
\begin{align}
  J_L &=  \frac{i}{2} \pqty{\dd{H} H^\dagger - H \dd{H^\dagger}}, & J_R &= - \frac{i}{2}  \pqty{H^\dagger \dd{H} - \dd{H^\dagger} H},
\end{align}
and we will be looking for solutions of the classical \ac{eom} at fixed values of the corresponding conserved charges
\begin{align}
  \mathcal{Q}_L &= \int \dd[3]{x} J_L^0, & \mathcal{Q}_R &= \int \dd[3]{x} J_R^0 .
\end{align}
Note that under the symmetry \(H \mapsto L H R^\dagger \), \(\mathcal{Q}_L\) and \(\mathcal{Q}_R\) transform as \(\mathcal{Q}_L \mapsto L \mathcal{Q}_L L^\dagger\), \(\mathcal{Q}_R \mapsto R \mathcal{Q}_R R^\dagger\), so the only invariant quantities are the eigenvalues of the charge matrices.
We will call these eigenvalues \(J_i\):
\begin{align}
  \spec(\mathcal{Q}_L) &= \set{J^L_1, J^L_2,  \dots, J^L_{N_F}} \\
  \spec(\mathcal{Q}_R) &= \set{J^R_1, J^R_2,  \dots, J^R_{N_F}}.
\end{align}

If for given fixed charges a solution exists which is homogeneous in space, this will be the solution of minimal energy in this sector. 
We make the following ansatz for such a homogeneous solution:
\begin{equation}\label{eq:ansatz}
	H_0(t) = e^{iM_Lt}Be^{-iM_Rt},
\end{equation}
where $M_{L,R}$ are in the Cartan subalgebra of $SU(N_F)$ are therefore related to the charges $\mathcal{Q}_{L,R}$, and $B$ is a self-adjoint $N_F\times N_F$ matrix~\cite{Loukas:2017lof,Banerjee:2019jpw}.

In a first step, we impose charge conservation:
\begin{align}
  \dot{\mathcal{Q}}_L &= -iVe^{iM_L t}\pqty{ \comm{M_L^2}{B B^\dagger} - 2 \comm{M_L}{BM_RB^\dagger}}e^{-iM_L t} = 0,\\
  \dot{\mathcal{Q}}_R &= iVe^{iM_R t}\pqty{ \comm{M_R^2}{B^\dagger B} -  2\comm{M_R}{B^\dagger M_LB}}e^{-iM_R t} = 0,
\end{align}
where $V=\text{Vol}(M_3)$.
From here, we find that $M_R$ either commutes or anti-commutes with $B$ and in either case, we can rewrite the Ansatz~\eqref{eq:ansatz} in the form
\begin{equation}\label{eq:GS}
  H_0 = e^{2i Mt}B,
\end{equation}
where $B$ is diagonal without loss of generality, as it can be diagonalized using the right action alone. We find
\begin{align}
  \mathcal{Q}_L &= -2V MB^2, & \mathcal{Q}_R &= 2VB^2M=-\mathcal{Q}_L .
\end{align}
The $\mathcal{Q}_{L,R}$ matrices are diagonal and have to be equal up to a sign.
This shows that not all possible charge configurations lead to a homogeneous ground state, as already observed for simpler models~\cite{Alvarez-Gaume:2016vff,Hellerman:2017efx,Hellerman:2018sjf,Banerjee:2019jpw}.

The other \ac{eom} takes the form
\begin{multline}
  \frac{\del}{\del t}\left[\frac{\del}{\del \dot H^*}\left( \Tr(\dot H^\dagger \dot H) \right)\right] + \frac{\del}{\del H^*}V(H, H^*) \\
  = \del^2_0 H  +2u (H^\dagger H)H + 2v\Tr(H^\dagger H) H+ \frac{R}{6}H = 0,
\end{multline}
and for our Ansatz it reads
\begin{equation}
  2 M^2 = u B^2 + v \Tr(B^2) + \frac{R}{12}.
\end{equation}
If we write the components of \(M\) and \(B\) as \(M_{ii} = \mu_i\) and \(B_{ii} = b_i\), the \ac{eom} take the equivalent form
\begin{align}
  2 \mu_i^2 &= u b_i^2 + v \sum_{k=1}^{N_F} b_i^2 + \frac{R}{12} , & i&=1, \dots, N_F
\end{align}
and the corresponding charges are
\begin{equation}
  J_i = \eval{\mathcal{Q}_L}_{ii} = 2 V b_i^2 \mu_i .
\end{equation}

\subsection{The energy of the ground state}\label{sec:groundstate}

In the following, for simplicity and ease of notation, we will choose to fix all the charges such that they are equal up to a sign, so \(M^2 = \mu^2 \Id_{N_F}\), \(B = b \Id_{N_F}\) with both \(\mu > 0\) and \(b > 0\).
Since $M$ is proportional to the charge matrix that lives in the algebra \(su(N)\), it must be traceless.
So it contains $N_F/2$ diagonal elements equal to $\mu$ and $N_F/2$ diagonal elements equal to $-\mu$.
The more general case is discussed in Appendix~\ref{sec:general}.               

If we choose all the charges to satisfy \(\abs{J_i} = J \), the \ac{eom} take the simple form
\begin{equation}
  2 \mu^2 = \pqty{u + v N_F } b^2  + \frac{R}{12} , 
\end{equation}
with the condition 
\begin{equation}
  J = 2 V b^2 \mu .
\end{equation}
It is convenient to assume \(J\) to be large and expand in series. The natural expansion parameter is \(\mathcal{J}\):
\begin{equation}
  \Jexp = J \frac{(u+v N_F)}{8 \pi^2} = 2 J \frac{ \alpha_h + \alpha_v}{N_F} = 2 J_{\text{tot}} \frac{\alpha_h + \alpha_v}{N_F^2},
\end{equation}
where \(J_{\text{tot}} = J N_F/2 \) is the total charge.
Then, \(\mu\) takes the form
\begin{equation}
  \mu = \pqty{\frac{2\pi^2}{V}}^{1/3}\Jexp^{1/3} + \frac{R}{72}\pqty{\frac{V}{2\pi^2}}^{1/3} \Jexp^{-1/3}+\order{\Jexp^{-5/3}}.
\end{equation}
Note, that the coefficient of the term \(\Jexp^{-1}\) happens to be zero on shell.
It is also convenient to define a charge density \(\rho = 2 \pi^2 \Jexp / V\) so that 
\begin{equation}
  \mu = \rho^{1/3} + \frac{R}{72} \rho^{-1/3}+\order{\rho^{-5/3}} .
\end{equation}

The expansion requires \(\mathcal{J} \gg 1\) and, observing that at the fixed point both \(\alpha_h\) and \(\alpha_v\) are of order \(\epsilon\), we see that the expansion is consistent in the regime
\begin{equation}
  J_{\text{tot}} \gg  \frac{N_F^2}{ \epsilon}.
\end{equation}
This is a typical feature of the large-charge expansion, where the total charge has to be the dominant large parameter in the problem. In the case at hand, also the number of \ac{dof} $N_F^2$ and the inverse coupling $1/\epsilon$ are large. In the three-dimensional vector model at the Wilson--Fisher point on the other hand, the only large parameter is the number of \ac{dof} $N$, so the condition on the charge is $J_{\text{tot}} \gg N$~\cite{Alvarez-Gaume:2019xxx}.
In the $O(2)$ model, there are no other large parameters, so $J\gg 1$~\cite{Hellerman:2015nra}. 
On a compact manifold, the fixed charge is associated to the scale $\rho\propto\Jexp/V$. Since there are no other dimensionful parameters in our problem, the energy scale \(\rho^{1/3}\) will control the tree-level and the quantum corrections to the energy of the ground state.

We can write the energy of the ground state as the Legendre transform of the Lagrangian.
We have
\begin{equation}
  \frac{E}{V} = \sum_{i=1}^{N_F} \mu_i \fdv{\mathcal{L}_H}{\mu_i} - \mathcal{L}_H = 4 \sum_{i=1}^{N_F} b_i^2 \mu_i^2 + u \sum_{i =1}^{N_F} b_i^4 + v \pqty{\sum_{i=1}^{N_F} b_i^2}^2 + \frac{R}{6} \sum_{i=1}^{N_F} b_i^2 .
\end{equation}
For our choice of charges, we obtain an expansion in \(\Jexp\), starting from \(\Jexp^{4/3}\):
\begin{equation}
  E = \frac{3}{2} \frac{N_F^2}{\alpha_h + \alpha_v} \pqty{\frac{2\pi}{V}}^{1/3} \bqty{ \Jexp^{4/3} + \frac{R}{36} \pqty{\frac{V}{2\pi^2}}^{2/3} \Jexp^{2/3} - \frac{1}{144} \pqty{\frac{R}{6}}^2 \pqty{\frac{V}{2 \pi^2}}^{4/3} \Jexp^0 + \order{\Jexp^{-2/3}}}
\end{equation}
or, in terms of the charge density \(\rho\),
\begin{equation}
  \label{eq:Ground-energy-rho}
  E =  \frac{3 V}{4 \pi^2} \frac{N_F^2}{\alpha_h + \alpha_v} \bqty{ \rho^{4/3} + \frac{R}{36} \rho^{2/3} - \frac{1}{144} \pqty{\frac{R}{6}}^2 \rho^0 + \order{\rho^{-2/3}} } .
\end{equation}
In fact, we could have predicted the power \(4/3\) of the leading term purely on dimensional grounds.
The energy density \(E/V\) has mass dimension \([E/V] = 4\), for systems with isolated fixed points, the curvature is not relevant in the \ac{rg} sense, and the only dimensionful parameter of the problem is the charge density which has dimensions \([\rho] = 3\).

Specializing the expressions to the case of a three-sphere of radius \(r_0\), in view of the state-operator correspondence, we have $V=2\pi^2 r_0^3$ and $R = 6/r_0^2$ and the energy is given by
\begin{equation}
  E = \frac{3}{2r_0} \frac{N_F^2}{\alpha_h + \alpha_v}  \bqty{ \Jexp^{4/3} + \frac{1}{6} \Jexp^{2/3} - \frac{1}{144}  \Jexp^0 + \order{\Jexp^{-2/3}}}.
\end{equation}

\subsection{Symmetry-breaking pattern}
\label{sec:symm-break-patt}

Working at fixed charge breaks the \(SU(N_F) \times SU(N_F) \times U(1)_B\) symmetry.
It is convenient to distinguish two effects.
If we expand a generic field as ground state plus fluctuations, the \(M\) matrix acts like a chemical potential and gives rise to a term which breaks the symmetries explicitly, while the \(B\) matrix is akin to a ground state that breaks the remaining symmetry spontaneously, as shown explicitly in Section~\ref{sec:Goldstone-spectrum-equal}.

When the ground state \(H_0 \) is written in the form of Eq.~(\ref{eq:GS}) it is clear that the explicit breaking only happens for the \(SU(N_F)_L\) symmetry, which is reduced to the commutant of \(M\), \emph{i.e.} the subgroup \(C(M) \subset SU(N_F) \times U(1)_B\) that commutes with \(M\).
Since \(B\) is proportional to the identity, the spontaneous breaking preserves a group \(C(M)\) embedded ``diagonally'' in \(C(M) \times SU(N_F) \times U(1)\), in the sense that \(B\) remains invariant under the adjoint action of \(C(M)\).
The full symmetry-breaking pattern is thus
\begin{equation}\label{eq:symmBrJeq}
  SU(N_F) \times SU(N_F) \times U(1) \overset{\text{exp.}}{\longrightarrow} C(M) \times SU(N_F)  \overset{\text{spont.}}{\longrightarrow}
   C(M) .
\end{equation}
By Goldstone's theorem, the low-energy dynamics is described by \(\dim(SU(N_F)) = N_F^2 - 1 \) \ac{dof}.

For clarity, we reorder the rows of the charge matrix such that it takes the form
\begin{equation}
  \mathcal{Q}_L = J
  \begin{pmatrix}[c|c]
    \Id & 0 \\ \hline
    0 & - \Id
  \end{pmatrix},
\end{equation}
where $\Id$ is the $N_F/2 \times N_F/2$ identity matrix.
Also, \(M \) takes the same form with a positive and a negative block, while \(B\) is still proportional to the identity:
\begin{align}
  M &= \mu {\begin{pmatrix}[c|c]
    \Id & 0 \\ \hline
    0 & - \Id
  \end{pmatrix}} , &
                     B &= b {\begin{pmatrix}[c|c]
                         \Id & 0 \\ \hline
                         0 &  \Id
                       \end{pmatrix}}.
\end{align}
It is now easy to see that the commutant of \(M\) is \(C(M) = SU(N_F/2) \times SU(N_F/2) \times U(1)^2\).
By construction, the Goldstone fields will arrange themselves into representations of this unbroken group.
We will discuss the precise form of the spectrum in Section~\ref{sec:Goldstone-spectrum-equal}.

\section{Decoupling of the fermions}
\label{sec:fermions}
Now that we have understood the effect of fixing the charge in the bosonic sector, we can discuss the fermionic and gauge sectors.
We expect the fermions to become massive, with the mass scale given by the charge density, in analogy to the case of the  three-dimensional supersymmetric theory with an isolated vacuum at large R-charge discussed in~\cite{Hellerman:2015nra}.
We will see in the following that this is indeed the case: all the fermions become massive, and this in turn decouples also the gluons.
The large-charge, low-energy physics is therefore described completely by the Goldstone fields that result from the symmetry breaking which we will analyze in detail in Section~\ref{sec:Goldstone-spectrum-equal}.

We start with the fermionic part of the Lagrangian~\eqref{eq:fullLag}, 
\begin{equation}
  \mathcal{L}_{\text{f}} = \Tr(\bar Q i\slashed{D} Q) + y \Tr(\bar Q_L H Q_R + \bar Q_R H^\dagger Q_L).
\end{equation}
Expanded around the ground state \(H = H_0(t)\) given in Eq.~\eqref{eq:GS}, the action takes the form
\begin{equation}
  \mathcal{L}_{\text{f, GS}} = \Tr(\bar Q i\slashed{D} Q) + y \Tr(\bar Q_L e^{2i M t} B Q_R + \bar Q_R B e^{-2i M t}  Q_L).
\end{equation}
It is convenient to redefine the fermionic fields \(Q\) to eliminate the time-dependent coupling and trade it for a mass term.
A possible choice is
\begin{align}
  \psi_L &= e^{-iM t} Q_L, & \psi_R = e^{iM t} Q_R,
\end{align}
so that the Lagrangian reads
\begin{equation}
  \label{eq:fermionic-action-psi}
  \begin{aligned}
    \mathcal{L}_{\text{f, GS}} &= \Tr(\bar\psi i\slashed{D} \psi) - \Tr( \bar \psi_L \gamma^0 M \psi_L) + \Tr( \bar \psi_R \gamma^0 M \psi_R) + y\Tr(\bar\psi B \psi ) \\
    &= \Tr(\bar \psi i\slashed{D} \psi) - \Tr( \bar \psi \gamma^0  \gamma^5 M \psi)  + y\Tr(\bar\psi B \psi ).
  \end{aligned}
\end{equation}
The simplest way to see if the fermions actually become massive and decouple is to write down the inverse propagator.
The zero-momentum limit of its determinant gives the product of the masses of the fields.
To see that, observe that in general the determinant is given by the product of the dispersion relations,
\begin{equation}
  \det( D^{-1}(\omega, p )) = \prod_{\text{fields}} (\omega^2 - m_f^2 - f_f(p^2)),
\end{equation}
where \(f_f(p)\) is a function that vanishes for \(p = 0\).
In a relativistic theory we expect \(f_f(p^2) = p^2\), but fixing the charge breaks Lorentz invariance and, as we will see in Section~\ref{sec:Goldstone-spectrum-equal}, in our case \(f(p^2)\) is a more general function that can be expanded in series for small \(p\).
By definition, \(m_f \) is the mass of the field and
\begin{equation}
  \eval{\det(D^{-1}(\omega, p))}_{\omega = 0, p = 0} = \prod_{\text{fields}} m_f^2.
\end{equation}
The inverse propagator corresponding to the action in Eq.~(\ref{eq:fermionic-action-psi}) is
\begin{equation}
  D^{-1}(\omega, p) = -\omega \gamma^0 + p_i \gamma^i - \gamma^0 \gamma^5 M + y B  ,
\end{equation}
so we just need to compute
\begin{equation}
  \prod_{f=1}^{4 N_F} m_f^2  = \det( D^{-1}(0, 0)) = \det( - \gamma^0 \gamma^5 M + y B).
\end{equation}
Using Dirac's representation of the gamma matrices, we have
\begin{align}
  \gamma^0 &=
        {\begin{pmatrix}
            1 & 0 \\
            0 & -1
          \end{pmatrix}} \otimes \Id_2 = \sigma_3 \otimes \Id_2  , &
                                                    \gamma^5 &=
                                                          {\begin{pmatrix}
                                                              0 & 1 \\
                                                              1 & 0
                                                            \end{pmatrix}} \otimes \Id_2 = \sigma_1 \otimes \Id_2,
\end{align}
so the determinant reads
\begin{equation}
    \det( - \gamma^0 \gamma^5 M + y B) = \det( \pqty{ - i \sigma_2 M + \Id_2 y B} \otimes \Id_2) 
  = \det( {\begin{matrix}[c|c]
      y B & -M \\ \hline
      M & y B
    \end{matrix}})^2 .
\end{equation}
Using the fact that 
\begin{equation}
  \det( {\begin{matrix}[c|c]
      A & B \\ \hline
      C & D
    \end{matrix}}) = \det( A - B D^{-1} C ) \det(D),
\end{equation}
and since $B$ and $M$ are diagonal matrices, we find
\begin{equation}
  \begin{aligned}
    \det( {\begin{matrix}[c|c]
        y B & -M \\ \hline
        M & y B
      \end{matrix}})^2 &= \prod_{i=1}^{N_F/2} \det(
    {\begin{matrix}
        y b_i + \frac{\mu_i^2}{y b_i} & 0\\
        0 & y b_i + \frac{\mu_i^2}{y b_i}
      \end{matrix}})^2 \det( {\begin{matrix}
        y b_i & 0 \\
        0 & y b_i 
      \end{matrix}})^2 \\
    &= \prod_{i = 1}^{N_F/2} \pqty{ \mu_i^2 + y^2 b_i^2 }^4 .
  \end{aligned}
\end{equation}
We see that both the Yukawa term (via the term \(y^2 b_i^2\)) and the kinetic term (via the term \(\mu_i^2\)) contribute to the final expression.
If all the charges are equal, also all the fermions have the same mass which is given by
\begin{equation}
  \begin{aligned}
    m_\psi &= \pqty{ \mu^2 + y^2 b^2}^{1/2} = \pqty{\frac{2 \pi^2}{V}}^{1/3} \pqty{ 1 + 2 \frac{N_F}{N_c} \frac{\alpha_y}{\alpha_h + \alpha_v}}^{1/2} \Jexp^{1/3} + \order{\Jexp^{-1/3}} \\
    &= \pqty{ 1 + 2 \frac{N_F}{N_c} \frac{\alpha_y}{\alpha_h + \alpha_v}}^{1/2} \rho^{1/3} + \order{\rho^{-1/3}}.
  \end{aligned}
\end{equation}
We see that all the fermion are massive, with a mass fixed by the charge density \(\rho\), and they decouple from the rest of the theory.

Once all the fermions of the theory have decoupled, at scales below $m_\psi$ the pure gauge sector starts running towards lower energies as pure Yang--Mills theory. The resulting theory gaps with an estimated confining scale $\Lambda_{YM}$ 
\begin{equation}
\Lambda_{YM} = m_{\psi} \exp\left[ - \frac{3}{22 \alpha_g (m_\psi)} \right] \ . 
\end{equation}
Here $\alpha_g (m_\psi)$ is very close to the \ac{uv} fixed-point value that is of order $\epsilon$.
The gluons will generically modify all the terms in the large-charge expansion of the energy in Eq.~(\ref{eq:Ground-energy-rho}).
These contributions are however exponentially suppressed as \(\order{e^{-1/\epsilon}}\) and can be neglected in our approximation.
Below the scale \(\Lambda_{YM}\) we have the Goldstone excitations that we will discuss in the following.

\section{Goldstone spectrum for equal charges}
\label{sec:Goldstone-spectrum-equal}

\subsection{Expansion to quadratic order}

We have seen in Section~\ref{sec:symm-break-patt} that the low-energy dynamics is described by \(N_F^2-1\) massless \ac{dof}.
To study their spectrum, we need to expand the fields at second order in the fluctuations \(\Phi\) around the ground state \(H_0(t)\):
\begin{equation}
  H(t,x) = \exp[ 2 i \mu t
    \begin{pmatrix}[c|c]
      \Id & 0 \\ \hline
      0 & - \Id
    \end{pmatrix}] \pqty{b \begin{pmatrix}[c|c]
    \Id & 0 \\ \hline
    0 &  \Id
  \end{pmatrix} + \Phi(t,x)}.
\end{equation}
Since we are only interested in the leading behavior at small momenta, we can neglect the term proportional to the curvature and write the Lagrangian as:
\begin{equation}\label{eq:quadrLag}
  \begin{aligned}
    L ={}& \Tr[ \del_\mu H^\dagger \del_\mu H ] - u \Tr[ H^\dagger H H^\dagger H] - v \Tr[H^\dagger H]^2 \\
    ={}&
    \Tr[\del_\mu \Phi^\dagger \del_\mu \Phi] -2 i \Tr[  \mu  (\Phi^\dagger \del_0 \Phi - \del_0 \Phi^\dagger \Phi)] + 4 \mu^2 \Tr[ \Phi^\dagger \Phi ]\\
    &- u \Tr[ (b + \Phi ) (b + \Phi )^\dagger (b + \Phi ) (b + \Phi )^\dagger ] - v \Tr[(b + \Phi ) (b + \Phi )^\dagger]^2 \\
    & + 4 \mu^2 b \Tr[\Phi + \Phi^\dagger] - 2 i  b \Tr[ \mu (\del_0 \Phi - \del_0 \Phi^\dagger)] + 4 \mu^2 b^2 N_F .
  \end{aligned}
\end{equation}
The terms in the last line vanish identically in the action, and we are left with an explicit mass term and the quartic potential.

The fluctuation \(\Phi(t,x )\) can always be expanded as
\begin{equation}
  \Phi(t,x) = \sum_{A = 0}^{N_F^2 - 1 } (h_A(t,x) + i p_A(t,x)) T^A,
\end{equation}
where \(T^0 = 1/\sqrt{2N_F} \Id\) and the \(T^a\) are the generators in the generalized Gell--Mann basis of \(SU(N_F)\), which satisfy the identity
\begin{equation}
	T_aT_b = \frac{1}{2}\left(\frac{\delta_{ab}}{N_F}+(d_{abk} + i f_{abk})T^k\right),
\end{equation}
where $f_{abc}$ are the structure constants and $d_{abc}$ the totally symmetric tensor of $SU(N_F)$.
The quadratic term is
\begin{equation}
  4 \mu^2 \Tr[\Phi^\dagger \Phi] = 2 \mu^2 \sum_A(h_A^2 + p_A^2)
\end{equation}
and in the notation of~\cite{Abel:2017ujy} it is written as a negative contribution to the mass:
\begin{align}
  m^2_{h_A h_B} &= - 4 \mu^2 \delta^{AB}, &   m^2_{p_A p_B} &= - 4 \mu^2 \delta^{AB}, &
   m^2_{p_A h_B} &= 0 .
\end{align}

The second-order expansion of the potential in terms of the \(h_A\) and \(p_A\) then takes the form
\begin{equation}
  V^{(2)} = \frac{1}{2} M^2_{h_A h_B} h_A h_B + \frac{1}{2} M^2_{p_A p_B} p_A p_B  + \frac{1}{2} M^2_{p_A h_B} p_A h_B, 
\end{equation}
where the \(M^2\) are those in~\cite{Abel:2017ujy} for the case of the background field state being \(H = b \Id =  \sqrt{2N_F} b T^0\).
Starting from (A.8) in~\cite{Abel:2017ujy}, we find
\begin{align}
  M^2_{h_0 h_0} &= m^2_{h_0 h_0 } + 6 b^2 \pqty{u + N_F v} = - 4 \mu^2 + 6 b^2 (u + N_F v ) = 8 \mu^2,\\
  M^2_{h_a h_b} &= m^2_{h_a h_b } + 2 b^2 \pqty{3u + N_F v}  \delta^{ab} = \pqty{-4 \mu^2 + 2 b^2 \pqty{3u + N_F v}}\nonumber\\
   &= 8 \mu^2 \frac{u}{u + N_F v} \delta^{ab} = 8 \mu^2 \frac{\alpha_h}{\alpha_h + \alpha_v} \delta^{ab},\\
  M^2_{p_A p_B} &= m^2_{p_A p_B } + 2 b^2 \pqty{u + N_F v} \delta^{AB} = \pqty{-4 \mu^2 + 2 b^2 \pqty{u + N_Fv}} \delta^{AB} = 0 ,\\
  M^2_{p_A h_B} &= m^2_{p_A h_B } = 0.
\end{align}

\subsection{Goldstone types and dispersion relations}

Let us pause a moment and see what we have found.
The effective mass matrix is diagonal.
Imposing the \ac{eom}, we find that there are \(N_F^2\) massless and \(N_F^2\) massive \ac{dof}.
This is not consistent with the symmetry breaking pattern \(C(M) \times SU(N_F) \to C(M)\) that we have discussed in Section~\ref{sec:symm-break-patt}.
The reason for this discrepancy is that we have decided not to consider the anomalous \(U(1)\) axial symmetry that is broken by quantum effects.
This means that one of the massless \ac{dof} found here is actually spurious, and we are left with the expected \(N_f^2 -1\).

At this point it would be natural to arrange them into the adjoint representation of \(SU(N)\) (plus the spurious singlet) since the fluctuations are literally written as linear combinations of the generators of the adjoint of \(SU(N)\) (plus an extra term proportional to the diagonal matrix \(T^0\)).
We need however to take into account that fixing the charge breaks both the Lorentz and the global \(SU(N)\) symmetries.
Because of the breaking of Lorentz invariance, we expect both relativistic (type I) and non-relativistic (type II) Goldstone bosons to appear~\cite{Nielsen:1975hm} and
a priori we do not know how our massless modes are organized into fields.
This information is encoded in the term linear in \(\mu\) with one time derivative appearing in Eq.~\eqref{eq:quadrLag}.
Remember that in our parametrization all the \(h_A\) are massive, while all the \(p_A\) do have a vanishing quadratic term \(M_{p_A p_B}^2\).
A type-II Goldstone can arise only if the fields \(p_A\) and \(p_B\) are related by a linear term proportional to \(\dot p_A p_B - p_A \dot p_B\).
In this case, the corresponding inverse propagator is
\begin{equation}
  \Delta^{-1}_{p_A p_B} =
  \frac{1}{2}\begin{pmatrix}
    \omega^2 - p^2  & 4 i \mu \omega \\
    -4 i \mu \omega & \omega^2 - p^2 
  \end{pmatrix} .
\end{equation}
Imposing \(\det(\Delta^{-1}) = 0\) and expanding for small momenta \(p\), we see that the pair $p_A,\,p_B$ corresponds to a type-II Goldstone and a massive mode with dispersion relations
\begin{align}
  \omega &= \frac{p^2}{4 \mu} + \dots, & \omega &= 4 \mu + \frac{p^2}{4 \mu} + \dots
\end{align}
How many such configurations are possible?
Consider the one-derivative term only for the \(p_A\) modes:
\begin{multline}
  \eval{  \Tr[ \mu (\Phi^\dagger \del_0 \Phi - \del_0 \Phi^\dagger \Phi)] }_{\Phi = p_A T^A} =  \sum_{A,B} \Tr[ \mu (p_A \dot p_B T^{A} T^{B} - \dot p_A p_B T^A T^B)]  \\
  =  \sum_{AB} \mu p_A \dot p_B \Tr[{ \begin{pmatrix}[c|c]
      \Id & 0 \\ \hline
      0 & - \Id
    \end{pmatrix} \comm{T^A}{T^B}} ] 
    = i \mu \sum_{ABC} p_A \dot p_B \hat f_{ABC} \Tr[{\begin{pmatrix}[c|c]
       \Id & 0 \\ \hline
       0 & - \Id
     \end{pmatrix} T^C }].
\end{multline}
How many such independent terms exist?
To answer this question, we need to look at the matrices that generate the Gell--Mann basis.
If \(T^C\) is off-diagonal, the trace vanishes identically.
As for the diagonal generators (the Cartan generators), the problem is equivalent to asking in how many ways we can write a traceless matrix with two entries where one entry is in the first \(N_F/2\) rows and one entry is in the last \(N_F/2\) rows.
There are clearly \((N_F/2)^2\) ways to do that.
This is also the number of type-II Goldstone bosons and each of them encodes two \ac{dof}.

The Goldstone spectrum is completed by \(N_F^2 - 1 - 2 \times N_F^2/4 = N_F^2/2 - 1  \) type-I Goldstones.
To compute their dispersion relations we need again to look at the linear-in-\(\mu\) term.
A cross term between a \(h_a\)  mode and a \(p_a \) mode gives a type-I Goldstone and a massive field.
There are two possibilities, since the mode \(h_0\) has a different mass from all the \(h_a\).
\begin{itemize}
\item For \(h_0\), the inverse propagator reads
  \begin{equation}
    \Delta^{-1}_{h_0 p} =
    \frac{1}{2}\begin{pmatrix}
      \omega^2 - p^2  - 8 \mu^2 & 4 i \mu \omega \\
      -4 i \mu \omega & \omega^2 - p^2 
    \end{pmatrix}.
  \end{equation}
  Then the massless and the massive mode have dispersion relations
  \begin{align}
    \omega &= \frac{p}{\sqrt{3}} + \dots & \omega &= 2 \sqrt{6} \mu + \frac{5p^2}{12 \sqrt{6}\mu}.
  \end{align}
  The presence of a linear mode with velocity \(1/\sqrt{3}\) was expected. This is a universal sector that appears for fixed charge in any scale-invariant theory.
In \(d+1\) dimensions, the tracelessness of the stress tensor for a free boson requires the low-energy action to be of the form \((\del_t \phi)^2 - 1/d (\nabla \phi)^2\).
\item For \(h_a\), the inverse propagator reads
  \begin{equation}
    \Delta^{-1}_{h_a p} =
    \frac{1}{2}\begin{pmatrix}
      \omega^2 - p^2  - 8 \frac{\alpha_h}{\alpha_h+ \alpha_v } \mu^2 & 4 i \mu \omega \\
      -4 i \mu \omega & \omega^2 - p^2 
    \end{pmatrix}.
  \end{equation}
  Then the massless and the massive mode have dispersion relations
  \begin{align}
    \omega &= \sqrt{\frac{\alpha_h}{3 \alpha_h + 2 \alpha_v}} p + \dots & \omega &= \sqrt{\frac{8 \pqty{3 \alpha_h + 2 \alpha_v}}{\alpha_h + \alpha_v}} \mu + \order{p^2}.
  \end{align}
  In this case the velocity is not fixed by scale invariance, but we have a constraint from causality \(0< \alpha_h/\pqty{3 \alpha_h + 2 \alpha_v} < 1\), which implies \(\alpha_h + \alpha_v > 0\). This constraint is satisfied at the fixed point since, using Eq.~\eqref{v1}, \(\alpha_h + \alpha_v = 0.6991 \alpha_h > 0\).  
\end{itemize}

How are the Goldstone fields organized into representations of the unbroken group \(C(M) = SU(N/2) \times SU(N/2) \times U(1)^2\)?
Once more we look at the term linear in \(\mu\).
The original \(SU(N)\) global symmetry is explicitly broken to \(SU(N/2) \times SU(N/2)\).
The \ac{dof} \(p_a\) in the adjoint of \(SU(N)\) decompose as the sum of two adjoints of $SU(N_F/2)$, a pair of bifundamentals (that together form a single bifundamental type-II Goldstone) and a singlet (the conformal Goldstone), as shown in Table~\ref{tab:Goldstone-spectrum}.

\begin{table}
  \centering
  \ytableausetup{centertableaux,smalltableaux}
  \begin{tabular}{L{4cm}cccc}
    \toprule
       type & I & I & I & II \\
    \ac{dof} & \(1\) & \(N_f^2/4 -1 \) & \(N_f^2/4 -1 \) &  \(2 \times N_f^2/4 \) \\
    velocity & \(1/\sqrt{3}\) & \(\sqrt{\frac{\alpha_h}{3 \alpha_h + 2 \alpha_v}}\) & \(\sqrt{\frac{\alpha_h}{3 \alpha_h + 2 \alpha_v}}\) & n/a\\[1em]
     \(SU(N_F/2) \times SU(N_F/2) \) representation & \((\mathbf{1}, \mathbf{1})\) & {  (
                                                                                                        \begin{ytableau}
                                                                                                          { } &{ }  \\
                                                                                                          { } \\
                                                                                                           \none[\scriptscriptstyle \vdots]\\
                                                                                                          { }
                                                                                                        \end{ytableau}, \(\mathbf{1}\) )} &
                                                                                                                                            {    (\(\mathbf{1}\),
                                                                                                                                            \begin{ytableau}
                                                                                                                                              { } &{ }  \\
                                                                                                                                              { } \\
                                                                                                                                              \none[\scriptscriptstyle\vdots]\\
                                                                                                                                              { }
                                                                                                                                            \end{ytableau})} &
                                                                      { (\ydiagram{1}, \ydiagram{1})} \\[1em]
    \bottomrule
  \end{tabular} 
  \caption{The Goldstone spectrum resulting from fixing the charges in the sector  \(\set{J, \dots, J, -J, \dots, -J}\). The \(N_F^2 - 1 \)
    DOF stemming from the breaking of the global symmetry are arranged into a singlet (the conformal Goldstone), two adjoints of \(SU(N_F/2)\) and a pair of bifundamentals (that together form a single bifundamental type-II Goldstone).
    The type-I Goldstones contribute to the zero-point energy according to their velocities.
    The type-II Goldstone has a quadratic dispersion relation and has zero velocity.
    Not represented here is the spurious singlet corresponding to the anomalous axial symmetry.
  }
  \label{tab:Goldstone-spectrum}
\end{table}

\subsection{Vacuum energy of the Goldstone fields}

Now that we have the Goldstone spectrum, we can compute the leading quantum correction to the energy formula, which is given by the zero-point energy of the type-I Goldstone bosons (since the type II Goldstones have no zero-point energy).

At low energy, the action for a Goldstone \(\phi\) with dispersion relation \(\omega = c p + \dots \) has the form
\begin{equation}
  S_G = \int\displaylimits_{\setR \times M_3} \dd{t} \dd{\mathbf{r}} \bqty{\frac{1}{2} (\del_t \phi)^2 + \frac{c^2}{2} (\nabla \phi)^2}.
\end{equation}
Its one-loop energy is then
\begin{equation}
  E_G =\frac{1}{2}\Tr(\log( - \del_t^2{} - c^2 \triangle)) = \frac{1}{4 \pi} \int_{-\infty}^{\infty} \dd{\omega} \sum_{\mathbf{p}} \log(\omega^2 + c^2 E(\mathbf{p})^2) ,
\end{equation}
where \(E(\mathbf{p})\) are the eigenvalues of the Laplacian on \(M_3\),
\begin{equation}
  \triangle f_{\mathbf{p}}(\mathbf{r}) + E(\mathbf{p})^2 f_{\mathbf{p}}(\mathbf{r}) = 0.
\end{equation}
The expression is clearly divergent, but we can use a zeta-function regularization.
If we  write \(\log(A) = - \eval{\dv{A^{-s}}{s}}_{s=0}\), we can integrate out the effect of the time:
\begin{multline}
  E_G = - \frac{1}{4 \pi} \eval{\dv{s}  \int_{-\infty}^{\infty} \dd{\omega}  \sum_{\mathbf{p}} \pqty{ \omega^2 + c^2 E(\mathbf{p})^2}^{-s}}_{s=0}  \\
  = -\frac{1}{2} \eval{ \dv{s} \sum_{\mathbf{p}} \pqty{ \frac{\Gamma(s-1/2)}{2 \sqrt{\pi} \Gamma(s)} c^{1-2s} E(\mathbf{p})^{1 - 2s }}}_{s=0} 
  = c \sum_{\mathbf{p}} E(\mathbf{p}).
\end{multline}
The sum over the energies can be identified with a special value of the zeta function \(\zeta(s | M_3)\) for the Laplacian on \(M_3\):
\begin{equation}
  E_G = c \sum_{\mathbf{p}} E(\mathbf{p}) = c \eval{\sum_{\mathbf{p}} E(\mathbf{p})^{-2s}}_{s=-1/2} = c \zeta(-1/2 | M_3).
\end{equation}
The velocities do not depend on the charges, so the contribution to the energy is necessarily of order \(\Jexp^0\).

Collecting all the Goldstones, we find that the total contribution to the energy is given by
\begin{equation}
  E_{0} = \frac{1}{2}\pqty{ 2 \times \pqty{ \frac{N_F^2}{4} - 1 } \sqrt{\frac{\alpha_h}{3 \alpha_h + 2 \alpha_v}} + \frac{1}{\sqrt{3}}} \zeta(-1/2|M_3) .
\end{equation}
Note that the \(N_F\)-scaling of \(E_0\) is the same that we had found for the tree-level term.
In the case of a sphere of radius \(r_0\), the zeta function is known and it takes the value~\cite{Elizalde:2012zza}
\begin{equation}
  \zeta(-1/2| S^3) = -\frac{0.414\dots}{r_0}.
\end{equation}
We can now write the full expression for the conformal dimension of the lowest state with charges \(\set{\underbrace{J, \dots, J}_{N_F/2}, \underbrace{-J, \dots, -J}_{N_F/2}}\):
\begin{multline}\label{eq:resultD}
  \Delta(J ) = r_0 E(S^3) = \frac{3}{2} \frac{N_F^2}{\alpha_h + \alpha_v}  \bqty{ \Jexp^{4/3} + \frac{1}{6} \Jexp^{2/3} - \frac{1}{144}  \Jexp^0 + \order{\Jexp^{-2/3}}} \\ - \pqty{ \pqty{ \frac{N_F^2}{2}  - 2} \sqrt{\frac{\alpha_h}{3 \alpha_h + 2 \alpha_v}} + \frac{1}{\sqrt{3}}} \times 0.212\dots
\end{multline}
This result has to be understood as a triple expansion in the three large parameters of the problem, \(1/\epsilon\), \(N_F^2 \) and \(\Jexp\).

The generic form of this expansion is
\begin{equation}
  \begin{aligned}
    \Delta(J) ={}& \frac{N_F^2}{\epsilon} \bqty{ \pqty{ c_{4/3} + \order{N_F^{-2}}} \Jexp^{4/3} + \pqty{ c_{2/3} + \order{N_F^{-2}}} \Jexp^{2/3} + \pqty{ c_{0} + \order{N_F^{-2}}}  + \order{\Jexp^{-2/3}}} \\
    &- \pqty{ \pqty{ \frac{N_F^2}{2}  - 2} d_1 + \frac{1}{\sqrt{3}}} \times 0.212 + \order{\Jexp^{-2/3}} + \order{\epsilon},
  \end{aligned}
  \end{equation}
where for each coefficient we have stressed the order of the expected \(1/N_F\) corrections and everything is understood up to order \(\order{\epsilon} \) corrections.

Some terms in the above expansion are universal and fixed completely by the symmetries of the problem and dimensional analysis:
\begin{itemize}
\item the energy density has dimension \(4\) and the charge density has dimension \(3\). This explains the leading \(\Jexp^{4/3}\) term;
\item the effect of the curvature, which is not relevant in the \ac{rg} sense, since the theory has no moduli space~\cite{Hellerman:2017veg}, is subleading. The Ricci curvature has dimension \(2\) and leads to an expansion in the dimensionless quantity \(R \rho^{-2/3}\);
\item the \(1/\sqrt{3}\) comes from the conformal Goldstone and is independent of \(\epsilon\) and \(N_F\);
\item the coefficient \(N_F^2/2 - 2\) is due to the symmetry breaking pattern and only depends on the choice of fixed charges.
\end{itemize}
Since here we started from a calculable \ac{cft}, which is described by a trustworthy linear sigma model, we were able to explicitly compute the coefficients \(c_i\) at leading order in \(N_F^2\) and \(1/\epsilon\). This is in contrast to earlier works where \acp{cft} which are not perturbatively accessible were studied at large charge and their Wilsonian couplings could not be determined within the framework of effective field theory.
In analogy to these cases, we expect the generic form of the charge dependence of the result for the conformal dimension~\eqref{eq:resultD} and its generic features to be common to a larger class of models with the same matter content and isolated fixed points, in which the fixed points are not perturbatively accessible.

\section{Conclusions and outlook}\label{sec:conclusions}

In this article we have studied an asymptotically safe \ac{cft} in four dimensions with gauge group $SU(N_C)$, $N_F$ fermions and an $N_F \times N_F$ complex matrix scalar field in a sector of large fixed global $SU(N_F) \times SU(N_F)$ charge compactified on $S^3$.
This is the first time that the large-charge limit has been considered in a non-supersymmetric relativistic \ac{cft} in four dimensions, and also the first time a non-supersymmetric \ac{cft} containing fermions has been considered at large charge.
At fixed large charge, the fermions receive large masses due to both the Yukawa term and the kinetic term %
and decouple from the dynamics.
This in turn decouples the gluons which are described by a confining Yang--Mills theory.
Finally we are left with the scalar sector which at fixed charge is governed by a time-dependent classical ground state and fluctuations around it.
The fluctuations are encoded by both relativistic and non-relativistic Goldstone bosons.
Analyzing the spectrum of these modes we find the expected conformal Goldstone with velocity \(1/\sqrt{3}\) and via a causality constraint for the other modes, a consistency condition for the couplings of the model \(\alpha_h + \alpha_v > 0\).

Our main result is the conformal dimension of the lowest-lying operator at large charge, which via the state-operator correspondence is given by the energy of the fixed-charge quantum corrected ground state.
This dimension is expressed in terms of an expansion in the large parameters $\Jexp$, $N_F$ and $1/\epsilon$, where the leading term in the large charge $\Jexp$ scales as $\Jexp^{4/3}$, as expected on dimensional grounds, with subleading terms that scale as $\Jexp^{2/3}$ and $\Jexp^{0}$.
A universal contribution at order $\Jexp^{0}$ comes from the Casimir energy of the relativistic Goldstones and depends only on the symmetry-breaking pattern.
Since we are dealing with a controlled \ac{cft}, we have complete control over all the coefficients appearing in our main result.

\medskip

There are number of further directions that can be explored.
On the one hand, one could push further the calculation of \ac{cft} data by studying three-point functions which, together with the conformal dimensions, encode the full theory in this limit.
Other operators in the same sector can also be understood in terms of excited states.
On the other hand, the methods that we have presented can also be applied to more general systems with fixed points in  non-perturbative regimes where $\epsilon$ is of order one.
In this case it would be necessary to introduce a non-linear sigma model description for the Goldstone bosons.
The large-charge approach is moreover useful for any \ac{cft} of the type discussed here, including four-dimensional systems with an interacting fixed point in the \acl{ir}.

\newpage

\subsection*{Acknowledgments}

D.O. acknowledges partial support by the \textsc{nccr 51nf40--141869} ``The Mathematics of Physics'' (Swiss\textsc{map}).
The work of S.R. is supported by the Swiss National Science Foundation under grant number \textsc{pp00p2\_183718/1}.
The work of F.S. is partially supported by the Danish National Research Foundation under grant number \textsc{dnrf:90}. 

\appendix
\section{Fixing general charges}\label{sec:general}

We can consider a more general charge configuration, where the homogeneous ground state takes the form
\begin{equation}
  H_0 = e^{2 i M t} B,
\end{equation}
with
\begin{align}
  M &= \pmqty{\dmat{\mu_1, -\mu_1, \mu_2, -\mu_2, \ddots }}, &  B &= \pmqty{\dmat{b_1, b_1, b_2, b_2, \ddots }}.
\end{align}
On this ansatz for the ground state, the \ac{eom} reduce to the algebraic equations
\begin{align}\label{eq:eom-alg}
 2 \mu_i^2 &= u b_i^2 + 2 v \sum_{k=1}^n b_k^2 + \frac{R}{12} , &  i &= 1, 2, \dots N_F/2
\end{align}
under the charge-fixing constraints
\begin{align}%
  \label{eq:charge-fix}
  \frac{J_i}{V} &= 2 b_i^2 \mu_i & i&= 1, 2, \dots, N_F/2 .
\end{align}
The energy of the configuration is then given by
\begin{equation}%
  \label{eq:EnergyGS}
  E = 4\sum_{k=1}^{N_F/2} J_k \mu_k + \frac{RV}{3} \sum_{k=1}^{N_F/2} b_k^2 + 4V v \pqty{ \sum_{k=1}^{N_F/2} b_k^2}^2 + 2 Vu \sum_{k=1}^{N_F/2} b_k^4 .
\end{equation}

We fix all the charges to be of the same order, \emph{i.e.} \(J_k =j_k J \) where \(j_k\) is order \(\order{1}\) and \(J\) is ``large'' and will be used as an expansion parameter.
Then we expand \(\mu_i\) as
\begin{equation}
  \mu_i = m_{1/3,i} J^{1/3} + m_{-1/3,i} J^{-1/3} + \dots
\end{equation}
and expand the \ac{eom} in powers of \(J\):
\begin{itemize}
\item at leading order (\(J^{2/3}\)) the \ac{eom} are
  \begin{equation}
    2 m_{1/3,i}^2 - \frac{u}{2V m_{1/3,i}} - \frac{v}{V} \sum_{k} \frac{j_k}{m_{1/3,k}} = 0
  \end{equation}
\item at next-to-leading order (\(J^{0}\)) we find
  \begin{equation}
    4 m_{1/3,i} m_{-1/3,i} - \frac{R}{12} + \frac{u}{2V} \frac{m_{-1/3,i}}{m_{1/3,i}^2} + \frac{v}{V} \sum_k  \frac{j_k m_{-1/3,k}}{m_{1/3,k}^2} = 0
  \end{equation}
\end{itemize}
and so on. The system can be solved order-by-order in \(J\) and then the energy can be consistently expanded.
We find
\begin{equation}
  E = c_{4/3} J^{4/3} + c_{2/3} J^{2/3} + \order{J^0},
\end{equation}
where
\begin{align}
  c_{4/3} &= \sum_k \pqty{4 j_k m_{1/3,k} + \frac{u}{2V} \pqty{\frac{j_k}{m_{1/3,k}}}^2} + \frac{v}{V} \pqty{\sum_k \frac{j_k}{m_{1/3,k}} }^2 ,\\
  c_{3/2} &= \sum_k\left( 4j_km_{-1/3,k} + \frac{R}{6} \frac{j_k}{m_{1/3,k}} - \frac{u}{V}\frac{j_k^2 m_{-1/3,k}}{m_{1/3,k}^3} -\frac{2v}{V}\frac{j_k}{m_{1/3,k}}\sum_i \frac{j_i m_{-1/3,i}}{m_{1/3,i}^2}
            \right).
\end{align}

\setstretch{1}

\printbibliography{}

\end{document}